\documentclass{elsart}
\usepackage{graphicx}
\begin{document}
\begin{frontmatter}
\title {Light Nuclei within  Nuclear Matter} 
\author[Rostock]{M. Beyer\thanksref{corresp}}
\author[unisa]{S.A.  Sofianos}
\author[Tokyo]{ N. Furutachi}
\author[Tokyo]{ S. Oryu}
% \email{sofiasa@science.unisa.ac.za}
%\affiliation{
\address[Rostock]{
              Institute of Physics, University of Rostock, 
               18051 Rostock,  Germany}
\address[unisa]{
Department of Physics, University of South Africa, 
               Pretoria 0003,  South Africa}
\address[Tokyo]{Department of Physics, Tokyo University of Science, 
               Noda, Chiba 278-8510, Japan}

\thanks[corresp]{Corresponding Author: michael.beyer@uni-rostock.de, tel. +49
  (381) 498 6773,
        fax. +49 (381) 498 6772}

\date{\today}% It is always \today, today,

\begin{abstract}
  We investigate the properties of $^3$He, $^4$He, $^6$He, $^7$Li and $^{16}$O
  nuclei in nuclear matter of finite temperature and density. A Dyson
  expansion of the many-body Green function leads to few-body equations that
  are solved using the Integro-Differential Equation Approach (IDEA) and the
  Antisymmetrized Molecular Dynamics (AMD) methods. The use of the latter
  method allows us to trace the individual movement of the wave packet for
  each nucleon and the formation and disintegration of quasi-nuclei in a
  changing thermodynamical nuclear matter environment.

\end{abstract}
\end{frontmatter}
%=========================================================
%\section{Introduction}

Nuclear fragments detected in a heavy ion collision at intermediate energies
play a central role to gain information on the properties and the dynamics of
nuclear matter under extreme conditions. One particular interesting example is
the liquid gas phase transition usually related to multi-fragmentation into
light nuclei~\cite{Feldmeier:1999je,Trautmann:2004cn}. For more and other
experiments see
e.g.~\cite{Borderie:1995xc,Borderie:1995sk,Nebauer:1998fy,INDRA00,Hudan:2002tn,Ghetti:2004uh}.
In practical calculations of heavy ion collisions~\cite{Bondorf:1995ua,Gross:1990cr,Danielewicz:1991dh,dan92,Stocker:1986ci,Kuhrts:2000zs,Colonna:2004zx,Feldmeier:1989st,Aichelin:1991xy,neb99,Lacroix:2004ne},
formation of clusters is described by kinematical (or geometrical) constraints
that allow certain nucleons to combine to larger clusters or larger clusters
to disintegrate. In some simulations these constraints are supplemented by a
probability to occur that is either related to some fit parameter or in more
microscopic approaches to the nucleon nucleon cross section. Depending on the
specific approach, e.g.  statistical multi-fragmentation
models~\cite{Bondorf:1995ua,Gross:1990cr}, microscopic transport models such
as Boltzmann
type~\cite{Danielewicz:1991dh,dan92,Stocker:1986ci,Kuhrts:2000zs,Colonna:2004zx},
quantum molecular dynamics~\cite{Feldmeier:1989st,Aichelin:1991xy,neb99}, or
recent more intermediate type of event generators~\cite{Lacroix:2004ne}, the
algorithms that deal with fragmentation are rather educated and powerful.
Despite the decisive role of fragmentation, the exact conditions under which
fragments are formed and which properties they have in such an excited nuclear
many-body system are little known. In addition, even for a particular heavy
ion collision at a given energy the formation of fragments could be at
different stages under quite different conditions.

The parameters that are introduced into the simulation to accommodate
fragmentation might be calculated microscopically. To motivate the objective
of our present work we remind the reader how the conditions of deuteron
formation are achieved in a microscopic statistical description of heavy ion
collisions such as the Boltzmann-Uehling-Uhlenbeck (BUU) approach. To this end
an overlap integral is introduced that cuts momentum components and hence
geometrically describes the effect of Pauli blocking. To be more specific,
formation of deuterons is only allowed if~\cite{Danielewicz:1991dh}
\begin{equation}
\int d^3q f\left(q+\frac{P_{\rm c.m.}}{2}\right)|\phi(q)|^2\leq F_{\rm cut}
\label{eqn:cut}
\end{equation}
holds, where $P_{\rm c.m.}$ denotes the c.m. momentum of the cluster and $q$
the relative momentum of the nucleons inside the wave function $\phi(q)$ of
the deuteron. The Pauli blocking is achieved by the nucleon momentum
distribution function $f$. The cut-off $F_{\rm cut}$ is a parameter that needs
to be fixed. Originally it has been treated as a fit parameter with a typical
value of $F^{\rm fit}_{\rm cut}\simeq0.2$. It has been shown that it is
possible to connect this parameter to a microscopic
calculation~\cite{Kuhrts:2000zs}.  Using in-medium deuteron wave functions
leads to $F_{\rm cut}\simeq0.15$ not quite the value used to fit to
experiments.  It could be that the higher value of $F^{\rm fit}_{\rm cut}$
simulates also evaporation from larger clusters that is not present in the
microscopic approach used in~\cite{Kuhrts:2000zs}.  The physical
interpretation of (\ref{eqn:cut}) is due to the Mott transition that occurs
when the binding energy of the cluster vanishes because of medium effects. For
an early paper on a microscopic calculation of the deuteron's Mott transition
see \cite{RMS,SRS}.  In the present work we proceed further along the pathway
of Ref.~\cite{Kuhrts:2000zs} investigating the Mott transition for larger
clusters up to $^{16}$O using rigorous few-body equations that include the
dominant medium effect, i.e. self energy corrections and Pauli blocking.  This
also extends previous investigations on the behavior of the nucleon deuteron
reaction~\cite{Kuhrts:2000zs,bey96,bey97,beyFB,kuhrts}, the
$^3$He/$^3$H~\cite{schadow} and the
$\alpha$-particle~\cite{sofia,Beyer:2003zz} along this lines.  These
investigations were based on the Alt-Grassberger-Sandhas (AGS) equations
\cite{AGS} which were modified to include the effects of the medium.

Because of the Pauli blocking the effective interaction between nucleons
becomes energy ($E$) dependent. To be specific we choose a temperature of
$T=10$ MeV that is characteristic for heavy ion collisions and also relevant
for neutron stars.  In the present work we first endeavor to transform the
$E$-dependent nucleon-nucleon interaction to an $E$-independent one using the
Marchenko inverse scattering method \cite{Mar}. This paves the way to go
beyond three- and four-nucleon systems described by integral equations in
momentum space by using configuration space formalisms. We employ here two
such formalisms, namely the IDEA \cite{IDEA1,Annals} and the AMD
\cite{ono,oryu} methods.  The former method is based on the assumption that
the dominant correlations in the medium are still of two-body nature and thus
one can obtain Faddeev-type equations for the corresponding two-body
amplitudes. The AMD method, on the other hand, enable us to study changes in
the density of the clusters embedded in the medium and obtain expectation
values $\langle r\rangle$ for the center of the wave packets for each nucleon
within the cluster.  In our studies we consider, the $^3$He, $^4$He, $^6$He,
$^7$Li and the $^{16}$O nuclei.

The Dyson expansion of the many-body Green function is used to obtain
thermodynamic cluster Green functions for the $A$-nucleon system embedded in
nuclear matter. The cluster Green functions are evaluated for an
uncorrelated medium. Hence, they can be formulated as resolvents and
respective resolvent equation can be derived for $A$ quasi-nucleons that can
be tackled with few-body techniques, however, modified effective potentials
and self energies (masses).  At a two body level this leads to the well known 
$t$-matrix equation~\cite{fet71},
\begin{equation}
    T_2(z) =   V_2 +  V_2 (1-f_1-f_2) R_0(z)  
                 T_2(z).
\label{T2}
\end{equation}
with $R_0(z)=(z-H_0)^{-1}$.The corresponding Schr\"odinger-type equation is
\begin{equation}
       (H_0  - z) \Psi(12)  +
             \sum_{1'2'} (1-f_1-f_2) V_2(12,1'2') \Psi(1'2') = 0.
\label{H0}
\end{equation}
The $f_i$ is the Fermi function given by
\begin{equation}
      f_i=\frac{1}{{\rm e}^{\beta(\varepsilon_i - \mu_{\mathrm{eff}})}+1}
\label{fE}
\end{equation}
with $\beta=1/k_BT$. We have used an effective mass approximation,
i.e. the quasi-particle energy  $\varepsilon$ is given by
\begin{equation}
     \varepsilon=\frac{k^2}{2m}+\Delta^{\rm HF}(k)
        \simeq\frac{k^2}{2m_{\mathrm{eff}}}+\Delta^{\rm HF}_0.
\label{vareps}
\end{equation}
The constant shift $\Delta^{\rm HF}_0$ can be absorbed in a redefinition of
the chemical potential $\mu_{\rm eff}=\mu-\Delta^{\rm HF}_0$.  For simplicity
we take the effective masses used in previous
calculations~\cite{Kuhrts:2000zs}.  Assuming that the two-body system is at
rest in the medium, then the kinetic energy term (\ref{H0}) is written as
\begin{equation}
     H_0= \frac{k_1^2}{2m_{\mathrm{eff}}}=\frac{k_2^2}{2m_{\mathrm{eff}}}
             =\frac{p^2}{2m_{\mathrm{eff}}}=\frac{1}{2}\, E
\label{H0eff}
\end{equation}
while the effective interaction that appears in (\ref{T2}) and (\ref{H0}) is
\begin{equation}
          V(E,r)=(1-2f(E))\;V_2(r), 
\label{Vef}
\end{equation}
where $E=z-E_{\mathrm{cont}}$ is the two-body center of mass  energy.
$E_{\mathrm{cont}}$ denotes the continuum edge as explained, 
e.g., in Ref.~\cite{RMS,SRS}.

To transform this potential to an equivalent $E$-independent one,
we employed   the Marchenko inverse
scattering method   \cite{Mar} which does not require to choose
 {\it a priori}  the shape and range of the potential.
The Schr\"odinger equation  is transformed 
to an integral one for the non-local function  $K(r,r')$,
\begin{equation}
        K_\ell(r,r')+{\mathcal F}_\ell(r,r')+\int_r^\infty
                K_\ell(r,s){\mathcal F}_\ell(s,r'){\rm d} s=0\,,
\label{March}
\end{equation}
where the kernel  ${\mathcal F}_\ell(r,r')$ is given by 
the Fourier transform of the  $S$-matrix,
\begin{eqnarray}
\nonumber
     {\mathcal F}_\ell(r,r')&=&\frac{1}{2\pi} \int_{-\infty}^{+\infty}
     h_\ell^{(+)}(kr)\left[1-S_\ell(k)\right]h^{(+)}_\ell(kr')
      {\rm d}k
\\
      &+&\sum_{n=1}^{N_b}
     A_{n \ell}h^{(+)}_\ell(b_n r)h^{(+)}_\ell(b_n r')\ .
\label{frr}
\end{eqnarray}
In the above equation, $h^{(+)}_\ell(z)$ is the Riccati-Hankel function
 and $A_{n \ell}$  are the asymptotic normalization  constants for 
the bound states $E_n=-\hbar^2/(2\mu)\, b_n^2$. The solution 
of Eq. (\ref{March}) can then provide us the underlying  
local $E$-independent  interaction  $ V(r)$. 
\begin{equation}
        V_\ell(r)=-2\;\frac{{\rm d} K_\ell(r,r)}{{\rm d}r}.
\label{Vr}
\end{equation}
More details and practical aspects of the method can be found in 
Ref. \cite{Mass}.  We only stress here that the Marchenko equation is 
fully equivalent to the Schr\"odinger equation. However, while in the 
latter equation one uses the knowledge of the potential from which
the physical information of the underlying system can
be extracted via the wave-function, in the Marchenko equation
the information at hand is the scattering and bound state data
embedded in the $S$-matrix.

An efficient way to study systems with $A\ge 3$ is to employ
 the IDEA method to the $A$-body. The method is  based on 
Hyperspherical Harmonics and the  $A$-body  wave function, due to 
pairwise acting forces, can be written as a sum of Faddeev-type 
amplitudes for the pairs $(i,j)$, 
\begin{equation}
       \Psi({\bf x})=\sum_{i<j\le A}\Phi({\bf r}_{ij},{\bf x})
      =H_{[L_m]}({\bf x}) \sum_{i<j\le A}    F({\bf r}_{ij},r),
\label{psiex}
\end{equation}
where $H_{[L_m]}({\bf x})$ is a harmonic polynomial of minimal degree $L_m$
and $r$ is the hyper-radius $r=(2/A)\sum_{i<j\le A} r^2_{ij}$. 
Using (\ref{psiex})  one  obtains the Faddeev-type equation for the amplitude 
$F({\bf r}_{ij},r)$ 
\begin{equation}
       \big (T-E\big)\,H_{[L_m]}({\bf x})F({\bf r}_{ij},r)=-V(r_{ij})
        H_{[L_m]}({\bf x})
 \sum_{k<l\le A}F({\bf r}_{kl},r).
\label{FanA}
\end{equation}
This equation may be modified by introducing in both sides 
the average of the potential over the unit hypersphere, {\em i.e}
the so-called hyper-central potential $V_0(r)$  \cite{IDEA1,Annals}
to obtain  an integro-differential equation for the 
(modified) Faddeev amplitude $\phi({\bf r}_{ij},r)$.
 Furthermore, assuming that the radial motion  and the orbital motion 
 are nearly decoupled, we may write
$$
     \phi({\bf r}_{ij},r)=P(z,r)/r^{(D-1)/2}\approx P_\lambda(z,r)
      u_\lambda(r)
$$
with $z=2 r^2_{ij}/r^2-1$ and  one obtains the  adiabatic 
approximation in which one extracts the potential 
surfaces $U_\lambda(r)$ that provide us with the binding energies 
and the scattering states of the system.

In contrast to the IDEA which is based on the Faddeev
decomposition of the wave function,  in the AMD method  
\cite{ono,oryu} one assumes that the antisymmetrized 
wave  function for the A-nucleons is given by a Slater determinant 
\begin{equation}
     |\Phi({\bf Z})\rangle=(1\pm P)\bigl({1}\big/{\sqrt{A!}}\bigr) 
           \det\big[ \varphi_{{\bf Z}_i}({\bf r}_j)  \big]
\label{wfA}
\end{equation}
constructed from the  the $i$-th nucleon wave function 
\begin{equation}
\label{wfone}
         \varphi_{{\bf Z}_i}({\bf r}_j)\equiv \langle {\bf r}_{j} |
         \varphi_{{\bf Z}_i}\rangle\,|\chi_i\rangle
       =\left(\frac{2\nu}{\pi}\right)^{3/4}
              \exp \left[-\nu({\bf r}_{j}-\frac{{\bf Z}_i}{\sqrt{\nu}})^2+
               \frac{1}{2}{\bf Z}^2_i\right]\,|\chi_i\rangle\,.
\end{equation}
In the above equations $P$ is the parity projection operator,
 ${\bf Z}$ is a set of (complex) parameters parameterizing the
nucleon, ${\bf Z} \,=\,({\bf Z}_1,{\bf Z}_2,{\bf Z}_3,...,{\bf Z}_A)$,
 $\nu$ is a width parameter for the wave packet and 
$|\chi_i\rangle=|\sigma_{i}\tau_{i}\rangle$ is the spin-isospin function 
for the nucleon $i$.
To obtain the solution one employs the time dependent 
variational principle 
\begin{equation}
        \delta \int_{t_1}^{t_2}{\rm d}t\frac{\langle\Phi({\bf Z})|\big[i\hbar
           \displaystyle\frac{{\rm d}}{{\rm d}t}-H\,\big]
          |\Phi({\bf Z})\rangle}
            {\langle\Phi({\bf Z})|\Phi({\bf Z})\rangle}=0 
\label{variata}
\end{equation}
where $H$ is the total Hamiltonian of the system \cite{ono,oryu}.

One of the main advantages of the AMD formalism is the
 possibility of tracing the momentum and  position of the 
individual wave packet. For this one 
defines the harmonic oscillator annihilation operator
\begin{equation}
            {\boldmath a}=\sqrt{\nu}{\bf r}_i
                      +\frac{i}{2\hbar\sqrt{\nu}}{\bf p}_i
\qquad {\rm with} \qquad
{\boldmath a}|\phi_{{\bf Z}_i}\rangle={\bf Z}_i|\phi_{{\bf Z}_i}\rangle\,.
\label{anop}
\end{equation}
where ${\bf Z}_i$ is a complex  vector with its real
and imaginary component are related to the average position and moment 
of the $i$-th wave packet,   
\begin{equation}
            {\bf Z}_i=\sqrt{\nu}{\bf D}_i
                      +\frac{i}{2\hbar\sqrt{\nu}}{\bf K}_i
\label{zrp}
\end{equation}
where
\begin{equation}
           {\bf D}_i=\frac{\langle\phi_{{\bf Z}_i}|{\bf r_i}|
          \phi_{{\bf Z}_i}\rangle}
           {\langle\phi_{{\bf Z}_i}|\phi_{{\bf Z}_i}\rangle}
\ \ \  {\rm and}\ \ \
     {\bf K}_i=\frac{\langle\phi_{{\bf Z}_i}|{\bf p}_i|
          \phi_{{\bf Z}_i}\rangle}
           {\langle\phi_{{\bf Z}_i}|\phi_{{\bf Z}_i}\rangle}\,.
\label{rpav}
\end{equation}
In our calculations we employ the general form of the nucleon-nucleon 
potential
\begin{equation}
   V(r)=(W+BP_{\sigma}-HP_{\tau}-MP_{\sigma}P_{\tau}) V(r)
%   +({\bf L{\cdot}S })V_{LS}(r) 
\end{equation}
\begin{table}[tb]
\centering{
\begin{tabular}{|l|c c c|}
\hline
\multicolumn{1}{|l|}{$\rho$ (fm$^{-3}$)}&
\multicolumn{1}{c}{$^{3}$He}&
\multicolumn{1}{c}{$^{4}$He}&
\multicolumn{1}{c|}{$^{16}$O}\\
\hline
%   rho      &     He3    &    He4      &  16O
 0           &  8.67     &   30.7    &  1630  \\
 0.003       &  3.84     &   18.1    &  1280  \\
 0.007       &  0.75     &   8.56     &  979.6   \\
 0.009       &   ---      &   5.70     &  874.8   \\
 0.017       &   ---      &      ---    &  628.0   \\
 0.034       &   ---      &   ---       &  507.8   \\
\hline
\end{tabular}
\caption{\label{ideabs}  Binding energies (in MeV) in the eaa
  for ${^3}$He,  $^{4}$He and  $^{16}$O
  systems for different densities $\rho$ (in  fm$^{-3}$)}
}
\end{table}
where the bare potential $V(r)$ is the Volkov \cite{Volkov} interaction which
is widely used in model nuclear structure calculations.  We first employ the
Wigner force only to study the influence of correlations on the bound states
of $^{3}$He, $^{4}$He and of $^{16}$O for a various densities using the IDEA
method. The binding energies obtained in the extreme adiabatic approximation
(eaa) are given in Table \ref{ideabs}. The $^3$He is the first to dissolve in
the medium while the $^{16}$O still is bound at the quite high density of
$\rho=0.034$\,fm$^{-3}$.  Note, however, that already the isolated $^{16}$O is
strongly over-bound in the eaa.  The disappearance of bound states is
attributed to the weakening of the attraction in the interaction with
increasing density $\rho$ of the medium.

\begin{figure}[t]
\includegraphics[width=7cm,angle=180]{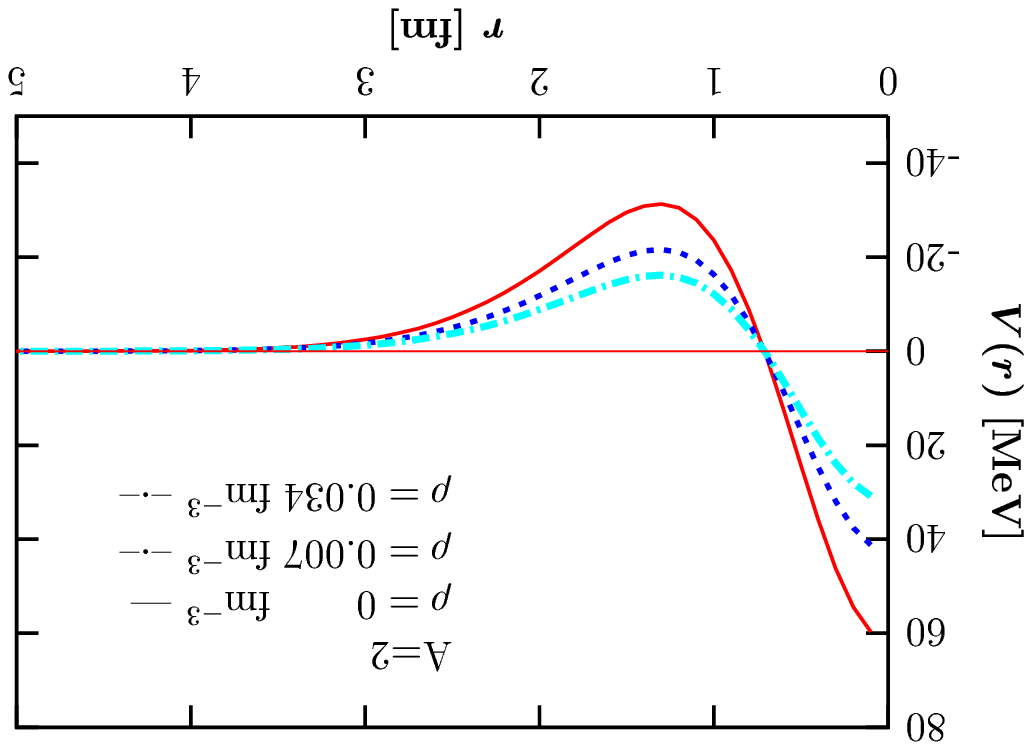}
\includegraphics[width=7cm,angle=180]{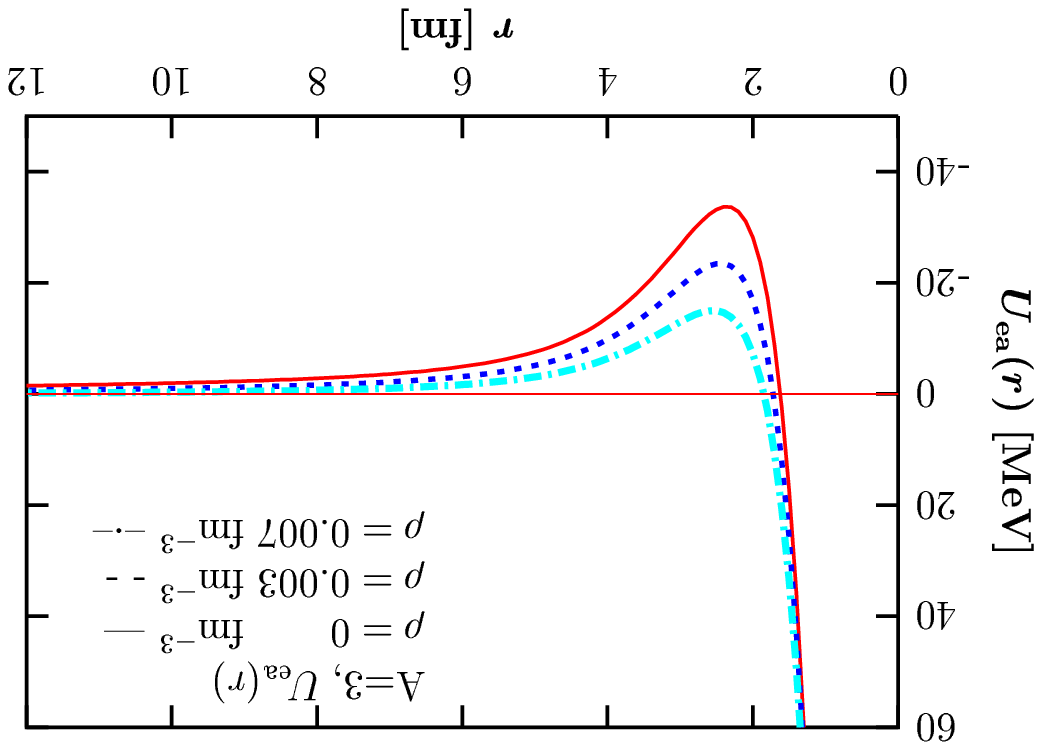}
\includegraphics[width=7cm,angle=180]{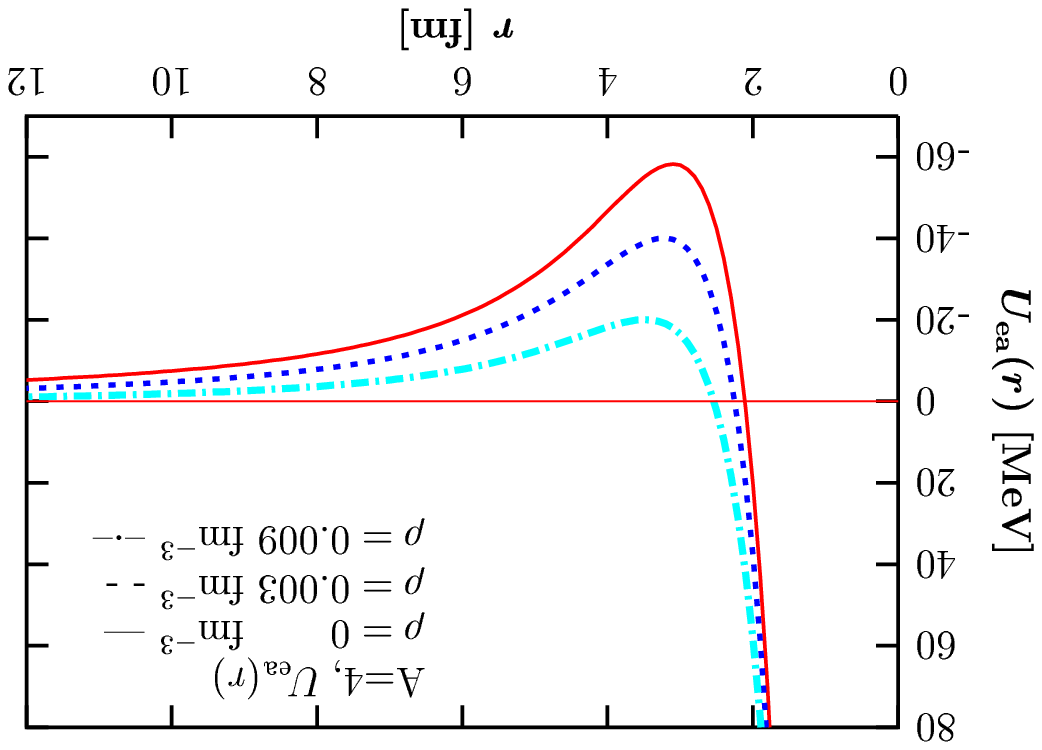}\hfill
\includegraphics[width=7cm,angle=180]{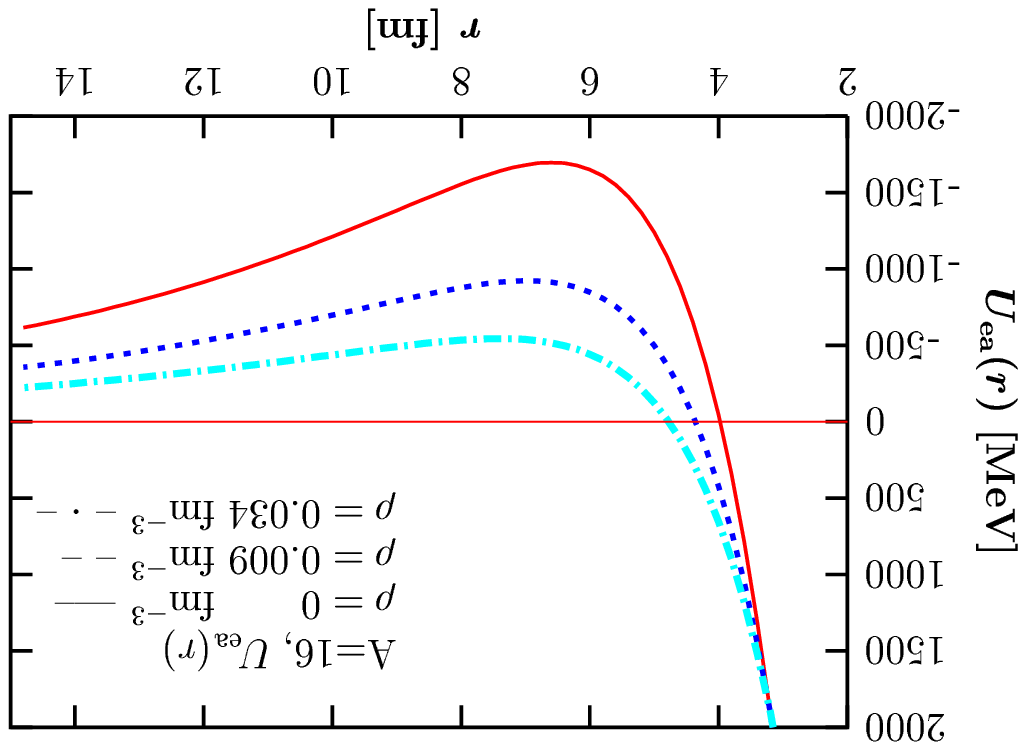}
\caption{\label{veaa}The two-body potential and the potential 
surfaces (in the extreme adiabatic approximation) for $A=3$,
$A=4$ and $A=16$ for different densities $\rho$.}
\end{figure}
\begin{figure}[b]
\centering
\includegraphics[width=6cm,angle=0]{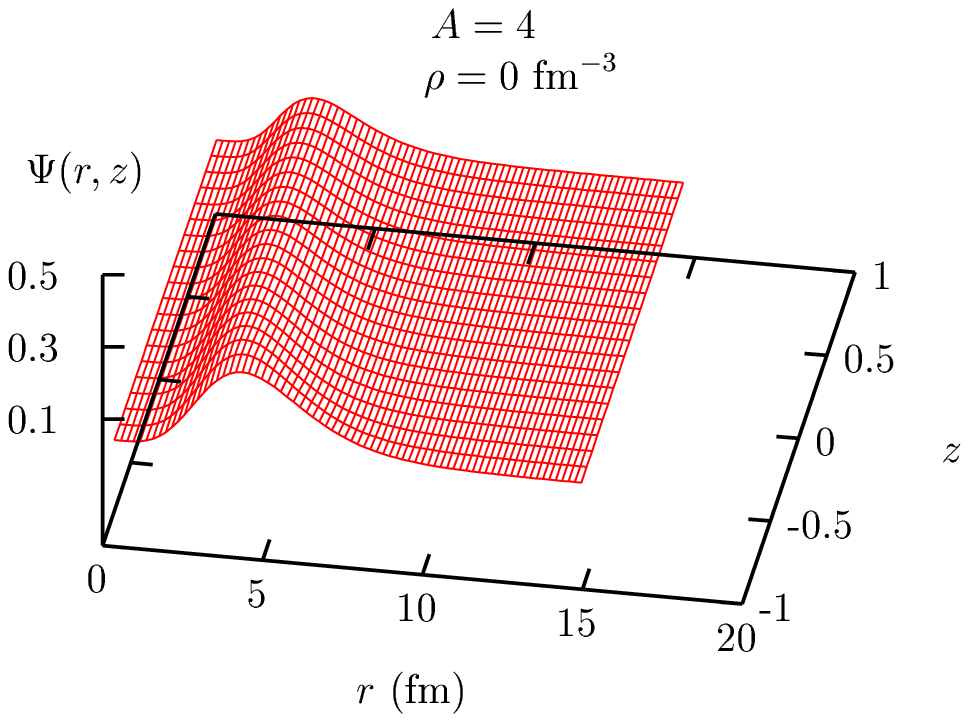}\hfill
\includegraphics[width=6cm,angle=0]{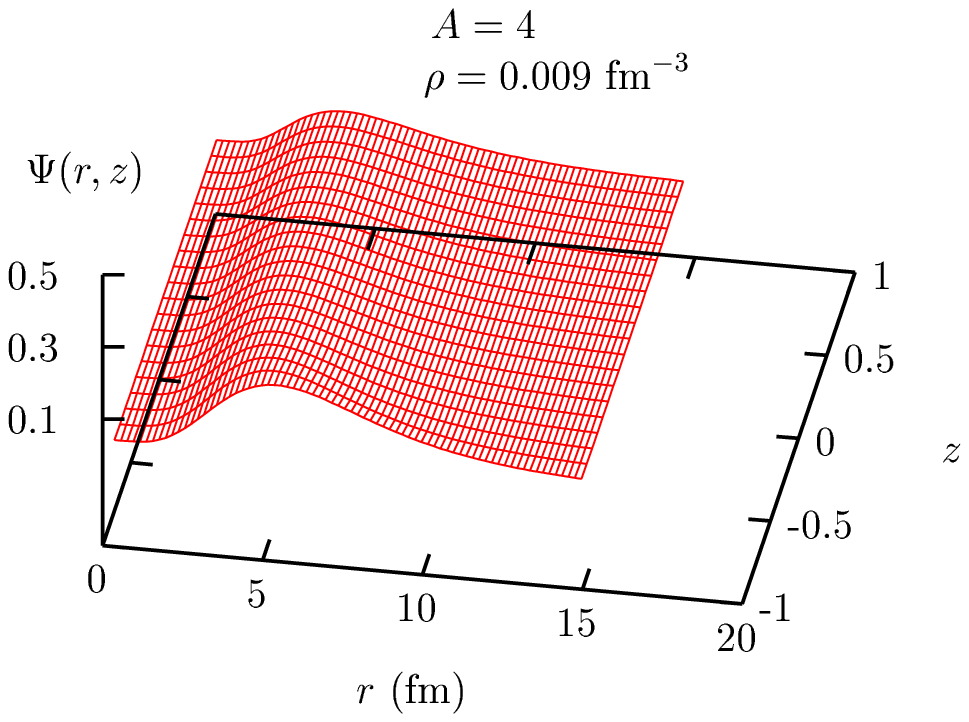}
\caption{\label{fadc}
The Faddeev components for the $A=4$ system for $\rho=0$\,fm$^{-3}$
and for for $\rho=0.009$\,fm$^{-3}$. The latter is just below the Mott
density.}
\end{figure}

This can be seen in Fig. \ref{veaa} where the potentials 
surfaces (in the eaa) are shown together with the two-body 
interaction. The weakening of the force can be seen also in 
Fig. \ref{fadc} where the Faddeev components $P(r,z)$ are plotted
for the $^4$He case. For $\rho=0.009$\,fm$^{-3}$ the component
is quite flat with the maximum shifted to $\sim5$\,fm
implying that the nucleons involved are already
quite far from each other. 

\begin{figure}[t]
\hspace*{-0.25cm}
\includegraphics[width=7cm,height=6.cm,angle=180]{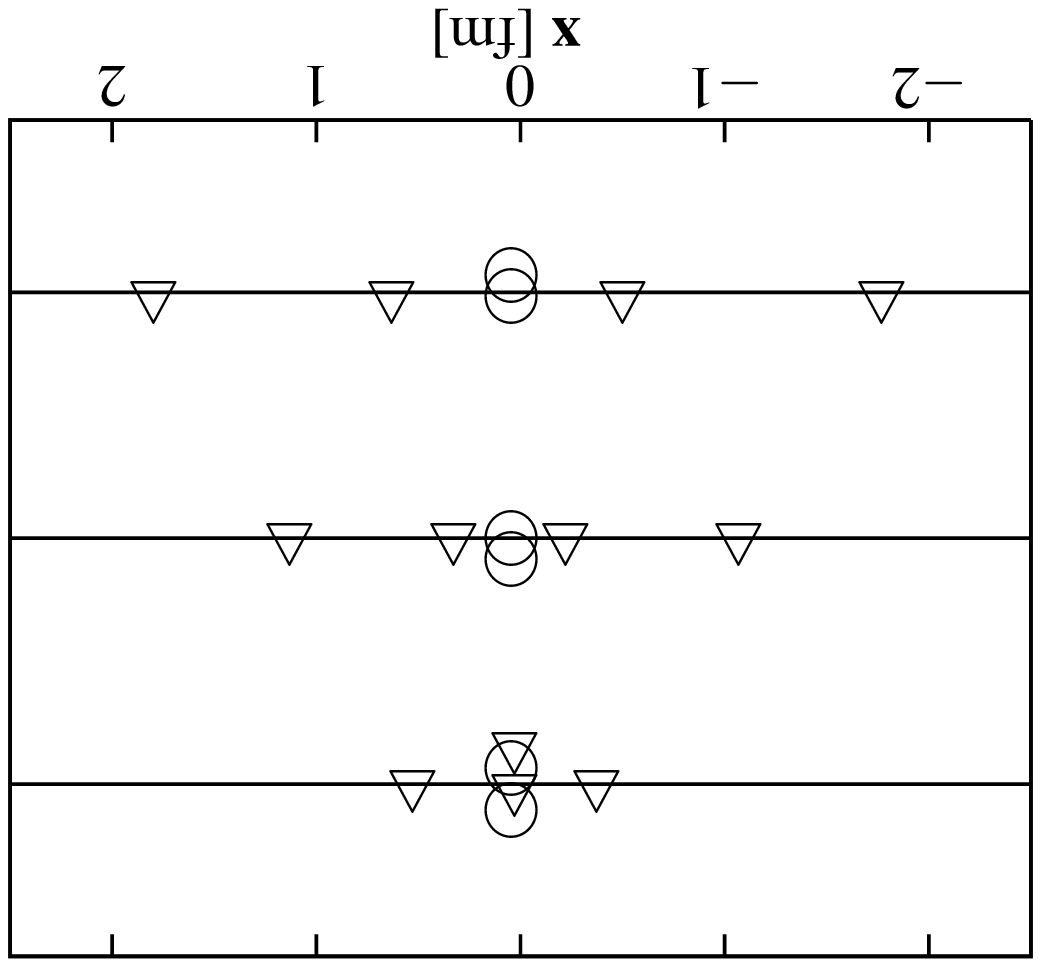} 
\hspace*{.1cm}
\includegraphics[width=7cm,height=6.0cm,angle=180]{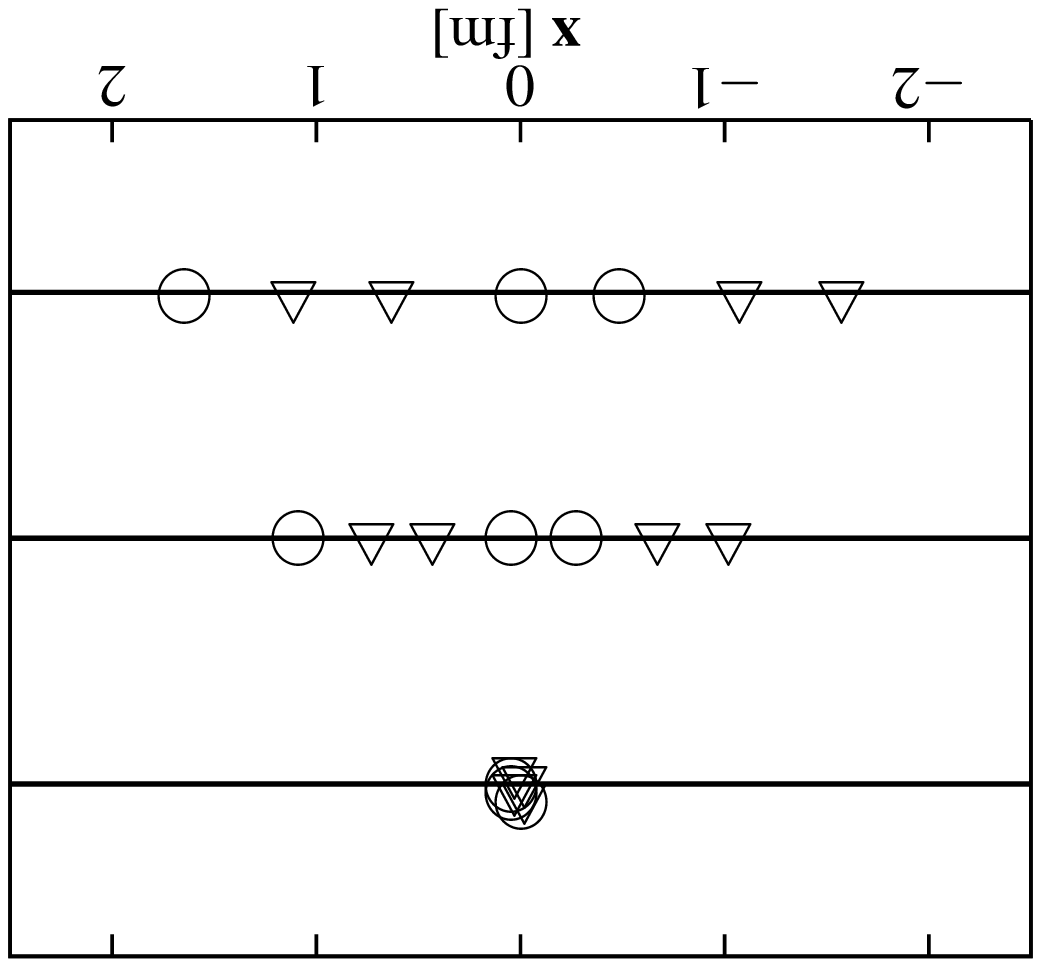}
 \caption{\label{fw6a7}
Expectation values $\langle r\rangle$ for the center of
the wave packets for  protons ({\large $\circ $}) and neutrons 
({\scriptsize $\triangle$})
for the $^{6}$He, left figure, and $^7$Li, right figure, 
with a Wigner force only. The three configurations (up to down)
correspond to $\rho=$0\,fm$^{-3}$, 0.009\,fm$^{-3}$ and 0.017\,fm$^{-3}$
respectively.
}
\end{figure}
\begin{figure}[t]
\centering{
\ \
\includegraphics[width=6.5cm,height=5.cm,angle=0]{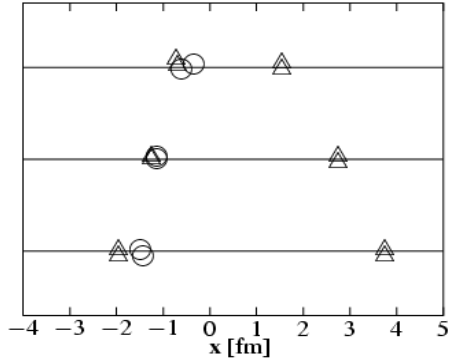}
\ \ 
}
 \caption{\label{fwm6}
Expectation values $\langle r\rangle$ for the center of
the wave packets for  protons ({\large $\circ $}) and neutrons 
({\scriptsize $\triangle$}) for the $^{6}$He
with Majorana force included. 
The three configurations (up to down)
correspond to $\rho=$0\,fm$^{-3}$, 0.009\,fm$^{-3}$ and 0.017\,fm$^{-3}$
respectively. }
\end{figure}

\begin{figure}[t]
\hspace*{-0.25cm}
\includegraphics[width=7cm,height=6.cm,angle=180]{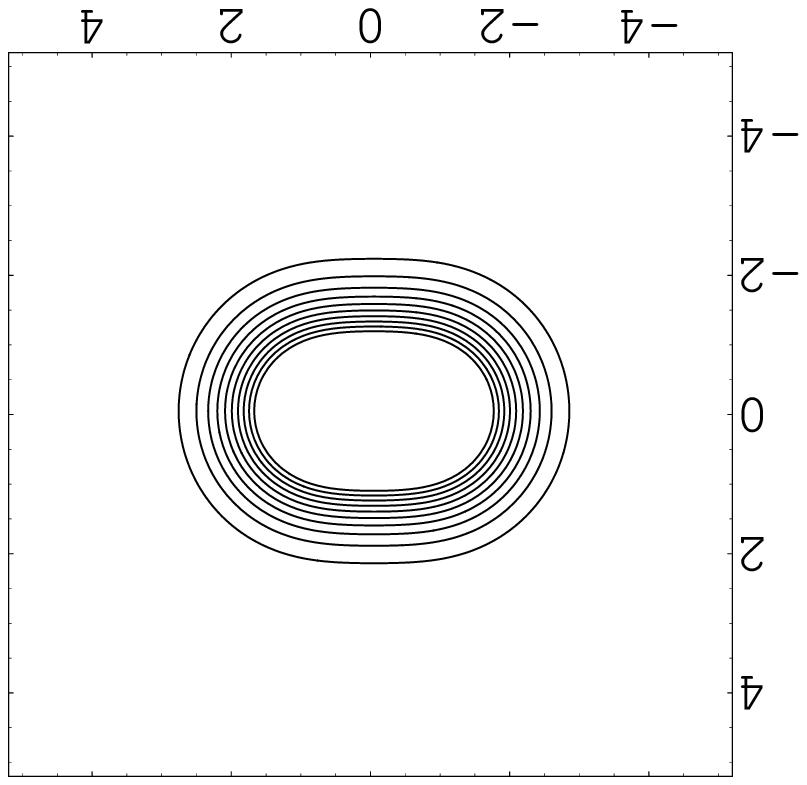} 
\hspace*{.1cm}
\includegraphics[width=7cm,height=6.cm,angle=180]{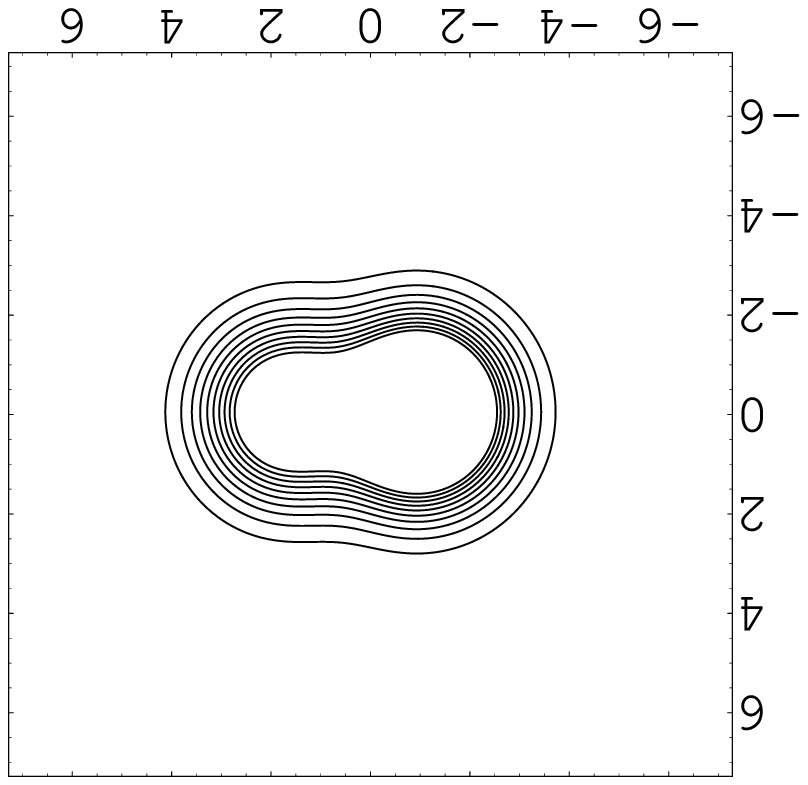} \\

\includegraphics[width=7cm,height=6.0cm,angle=180]{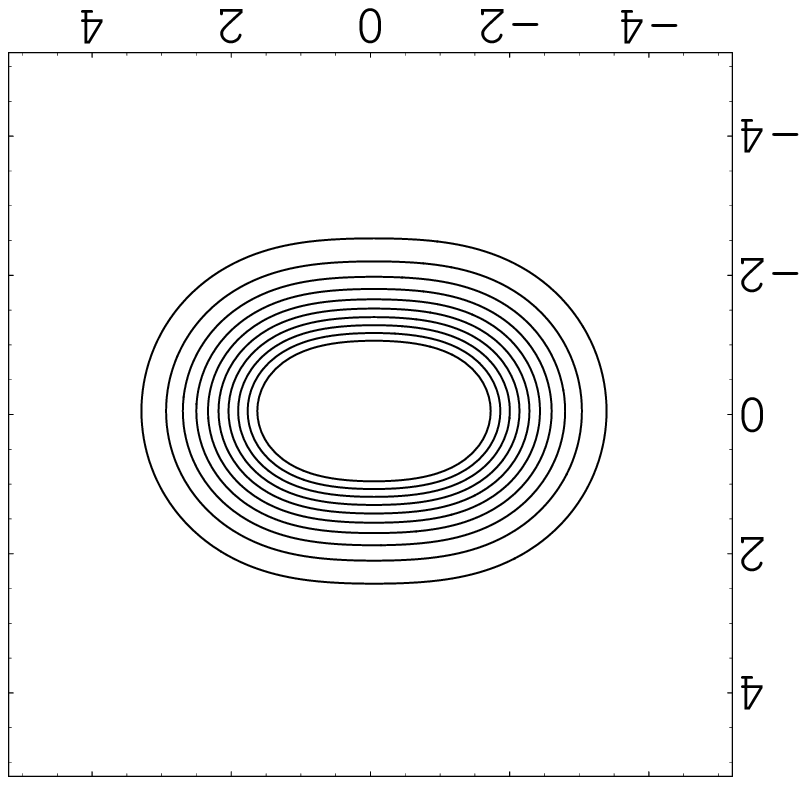}
\hspace*{.1cm}
\includegraphics[width=7cm,height=6.cm,angle=180]{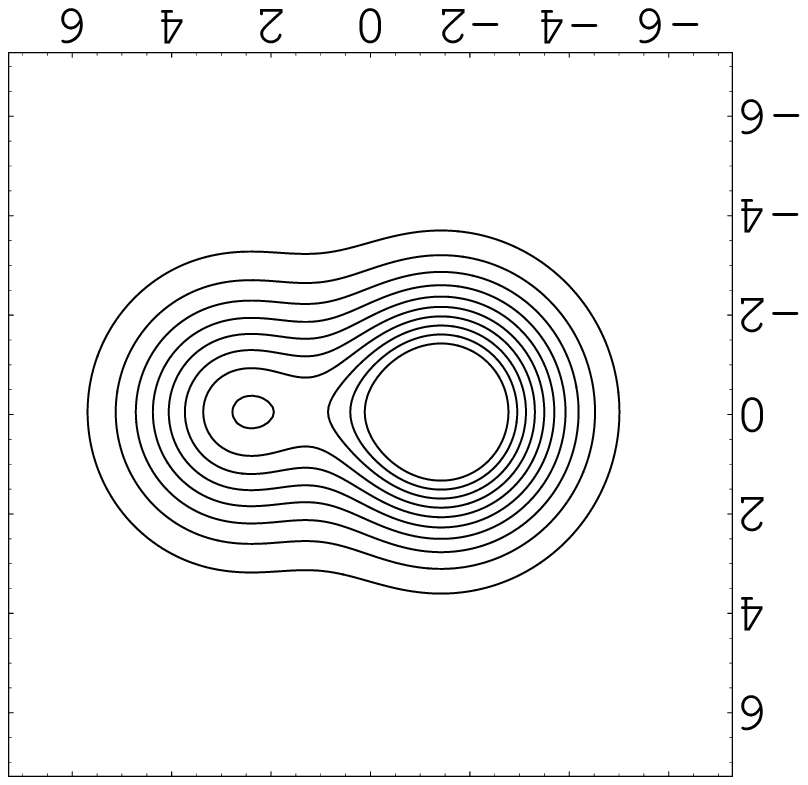} \\
\includegraphics[width=7cm,height=6.cm,angle=180]{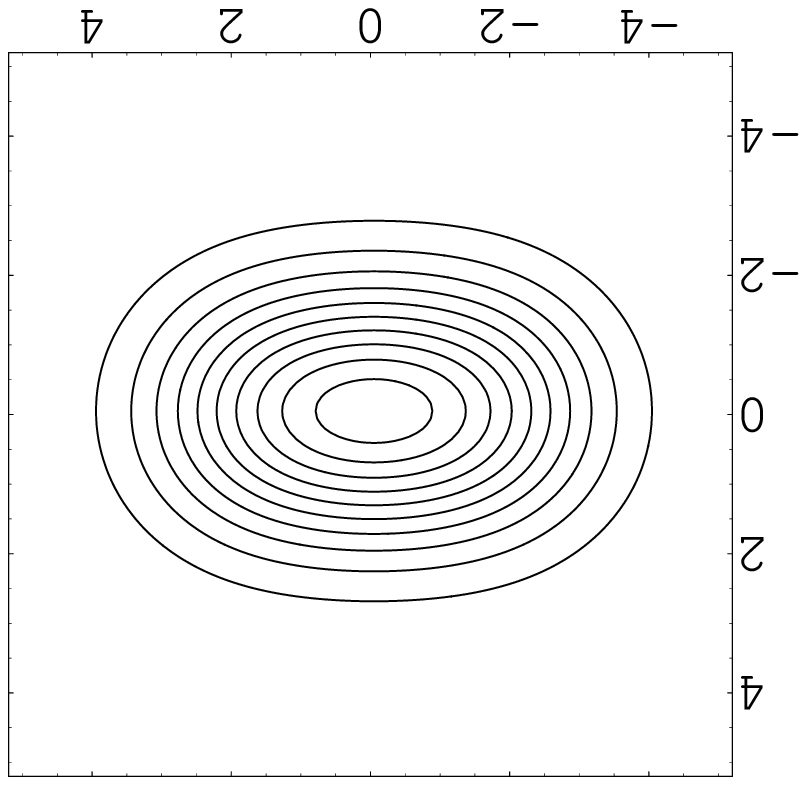} 
\hspace*{.1cm}
\includegraphics[width=7cm,height=6.cm,angle=180]{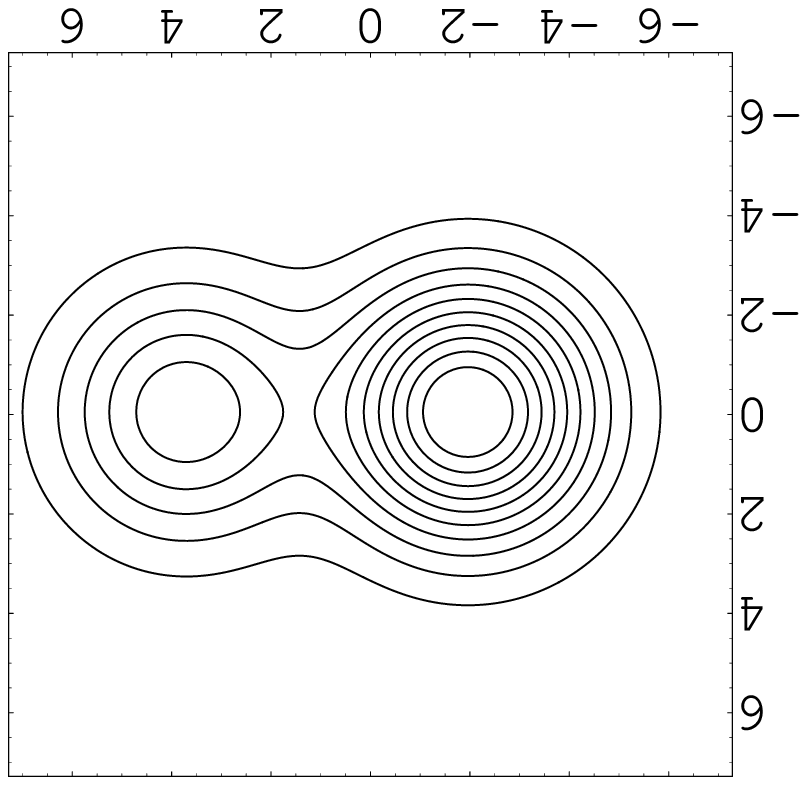} 
 \caption{\label{fig5}
Two-dimensional density plots for the $^{6}$He system with a 
Wigner force (left panel) and
with Majorana force included (right panel) for $\rho=0$\,fm$^{-3}$,
$0.007$\,fm$^{-3}$ and $0.017$\,fm$^{-3}$ (top to bottom).
The figure scales are in fm.
In the left panel the outermost line corresponds to the value
of 0.01\, fm$^{-2}$ and gradually increases by 0.5\, fm$^{-2}$
until it reaches the value of  5\, fm$^{-2}$ for the inermost line.
 In the right panel the correponding numbers are 0.01\, fm$^{-2}$ until
2\, fm$^{-2}$ in steps of .2\, fm$^{-2}$.
}
\end{figure}

As already mentioned, the main advantage of the AMD method is
the  possibility of tracing  the movement of an
 individual proton or neutron by calculating the
expectation value $\langle r\rangle$ for the
corresponding  wave packet.
In Fig. \ref{fw6a7} we show the  results for the $^6$He and $^7$Li
clusters with the Wigner force only. In the $^6$He case the 
two neutrons are squeezed out of the cluster
leaving the $\alpha$-particle behind which, however, also
is enlarged and ready to fall apart. The  $^7$Li is at first 
quite compact but  two neutrons and a proton start to escape 
with increasing $\rho$ and already at $\rho=0.017$\,fm$^{-3}$ 
the 7 particles are not bound. It is interesting to note
that although the $\alpha$ particle appears to be in the middle,
the other particles always arranged themselves in an 
$\alpha$ particle formation. 

The effects of the inclusion of the Majorana force in the 
calculation is shown in Fig. \ref{fwm6}. The differences
generated by the inclusion of this force
is  striking.  The $\alpha$ particle acts as an attractive center
for the two neutrons forming a halo nucleus which 
are escaping with increasing $\rho$. 
In the case where only a Wigner force is considered,
 the two neutrons are escaping in either side of the 
$\alpha$ particle. However, the inclusion of spin-isospin component 
in the nuclear force generates two sub-clusters consisting
of an $\alpha$ cluster and the two neutrons which tend to split 
first prior an overall disintegration occurs.
These differences are 
best seen on the two--dimensional  Fig. \ref{fig5}
obtained by integrating over the $z$-axis.

Our conclusions can be summarized as follows. 
i) 
The dynamical behavior of a nucleonic cluster
in a medium can be reliably studied using a combination
of methods namely, a cluster mean field approximation of the
Dyson Green function expansion,  an inverse scattering
method for transforming the $E$-dependent potential to 
an $E$-independent one and few-body methods.
 ii)
The construction of the $E$-independent interactions
paves the way to study the behaviour of light nuclei (A$\sim$3-20)
in a medium using configuration space formalisms. 
iii) 
The use of the AMD method allows us to trace
the individual movement of the nucleonic
wave packet and the formation of sub-clusters 
when the density increases. This feature is particularly
suited to explore correlations as has been done, e.g., for
light clusters in a recent experiment~\cite{Ghetti:2004uh}, and also get
insight into a possible evaporation mechanism of light clusters.
iv) 
The use of Wigner forces alone seems to be insufficient for a 
proper description of the clusters while the inclusion of the
Majorana force generated dramatic changes in the clustering
picture of the particles. However, in order to obtain a complete
picture on the behavior of quasi-nuclei in a changing nuclear
matter environment, more realistic forces that include also the
Bartlett and Heisenberg components as well as an LS-force
should be employed. Although the effects of the former two 
components are expected to be small, 
 the role played by the
latter force in these extremely loose bound systems
could be quite important
and therefore its inclusion  in studies concerning 
disintegration of clusters is warranted.

\section*{Acknowledgment}
Financial support from the FRCCS of Tokyo University  of Science
to one of the author (SAS) is greatly acknowledged.
\clearpage

%%%%%%%%%%%%%%%%%%%%%%%%%     REFERENCES     %%%%%%%%%%%%%%%

\end{document}